\documentclass[11pt,onecolumn]{IEEEtran}

\usepackage[utf8]{inputenc} 
\usepackage[T1]{fontenc}
\usepackage{url}
\usepackage{ifthen}
\usepackage{cite}
\usepackage[cmex10]{amsmath} 
\usepackage{graphicx}
\usepackage{amsthm,amssymb,amsfonts}
\usepackage{xcolor}
\usepackage{mathtools, cuted}
\usepackage{caption}
\usepackage{subfigure}

\newcommand{\bit}{\begin{itemize}}
	\newcommand{\eit}{\end{itemize}}
\newcommand{\bcor}{\begin{cor}}
	\newcommand{\ecor}{\end{cor}}
\newcommand{\beq}{\begin{equation}}
\newcommand{\eeq}{\end{equation}}
\newcommand{\beqn}{\begin{equation}}
\newcommand{\eeqn}{\end{equation}}
\newcommand{\bea}{\begin{eqnarray}}
\newcommand{\eea}{\end{eqnarray}}
\newcommand{\bean}{\begin{eqnarray*}}
	\newcommand{\eean}{\end{eqnarray*}}
\newcommand{\ben}{\begin{enumerate}}
	\newcommand{\een}{\end{enumerate}}
\newcommand{\bdefn}{\begin{defn}}
	\newcommand{\edefn}{\end{defn}}
\newcommand{\bnote}{\begin{note}}
	\newcommand{\enote}{\end{note}}
\newcommand{\bprop}{\begin{prop}}
	\newcommand{\eprop}{\end{prop}}
\newcommand{\blem}{\begin{lem}}
	\newcommand{\elem}{\end{lem}}
\newcommand{\bthm}{\begin{thm}}
	\newcommand{\ethm}{\end{thm}}
\newcommand{\bconj}{\begin{conj}}
	\newcommand{\econj}{\end{conj}}
\newcommand{\bconstr}{\begin{constr}}
	\newcommand{\econstr}{\end{constr}}
\newcommand{\bpf}{\begin{proof}}
	\newcommand{\epf}{\end{proof}}
\newcommand{\bc}{\begin{center}}
	\newcommand{\ec}{\end{center}}
\newcommand{\bprf}{{\em Proof: }}
\newcommand{\eprf}{\hfill $\Box$}

\newtheorem{thm}{Theorem}[section]

\newtheorem{lem}[thm]{Lemma}
\newtheorem{cor}[thm]{Corollary}
\newtheorem{defn}{Definition}[section]
\newtheorem{note}{Remark}[section]

\newcommand{\calc}{\mbox{$\mathcal{C}$}}

\IEEEoverridecommandlockouts
\begin{document}
\title{On the Performance Analysis of Streaming Codes over the Gilbert-Elliott Channel}
 \author{
   \IEEEauthorblockN{Myna Vajha, Vinayak Ramkumar, Mayank Jhamtani, P. Vijay Kumar\ \\}
   \IEEEauthorblockA{Department of Electrical Communication Engineering, Indian Institute of Science, Bangalore. \\
   	email: \{mynaramana, vinram93, mayankjhamtani26, pvk1729\}@gmail.com}
}

\maketitle

\begin{abstract}
The Gilbert-Elliot (GE) channel is a commonly-accepted model for packet erasures in networks. Streaming codes are a class of packet-level erasure codes designed to provide reliable communication over the GE channel. The design of a streaming code may be viewed as a two-step process. In the first, a more tractable, delay-constrained sliding window (DCSW) channel model is considered as a proxy to the GE channel. The streaming code is then designed to reliably recover from all erasures introduced by the DCSW channel model.  Simulation is typically used to evaluate the performance of the streaming code over the original GE channel, as analytic performance evaluation is challenging.
 In the present paper, we take an important first step towards analytical performance evaluation.  Recognizing that most, efficient constructions of a streaming code are based on the diagonal embedding  or horizontal embedding  of scalar block codes within a packet stream, this paper provides upper and lower bounds on the block-erasure probability of the underlying scalar block code when operated over the GE channel. 

\end{abstract}


\section{Introduction}

Achieving ultra-reliable, low-latency communication is a key cornerstone of the 5th generation (5G) cellular systems as it enables next-generation applications such as augmented reality, assisted driving and telesurgery.  Packet erasure, i.e., packet drops, are commonplace in a communication network and erasure coding represents a resource-efficient means of tackling them. The Gilbert-Elliott (GE) \cite{gilbert} \cite{elliott} channel is commonly used to model the packet erasures that take place over networks \cite{HasHoh,HohGeiHas,MalMedYeh}.  This motivates the design of efficient packet-level erasure codes for GE channel that are designed to be decoded under a strict decoding-delay constraint.  
However, the direct design of decoding-delay-constrained erasure codes with guaranteed performance over the GE channel is challenging.  To circumvent this, an approach that has been followed in the literature on streaming codes, is to first identify a delay-constrained, sliding-window (DCSW) channel model 
\cite{BadrPatilKhistiTIT17} that approximates the GE channel at hand, and then design a streaming code that is effective when operated over the DCSW channel. 
The DCSW channel model is characterized by four parameters.  In an $(a,b,w,\tau)$ DCSW channel, within any sliding window of length $w$ there can be either $\leq a$ random erasures or else a burst erasure of size $\leq b$.  In addition, a $\tau$-packet decoding-delay constraint is imposed.  Thus, a streaming code that claims to provide reliable communication over the $(a,b,w,\tau )$ DCSW channel, is required to correctly recover the $t$-th packet by the time the $(t+\tau)$-th packet arrives, provided that the erasure pattern encountered is compatible with the $(a,b,w,\tau )$ DCSW channel model. This channel is non-trivial only for $a \le b$.  It was shown in \cite{BadrPatilKhistiTIT17}, \cite{NikDeepPVK} that without loss of any generality, the parameter $w$ can be set equal to $\tau+1$.  As a result, three parameters $\{a,b,\tau\}$ suffice to characterize the DCSW channel, and we will speak of an $(a,b,w,\tau)$ DCSW channel model, as an $(a,b,\tau)$ DCSW channel model, with the understanding that $w=\tau+1$.  The $(a,b,\tau)$ DCSW channel is a generalization of an earlier burst-only channel model considered in \cite{MartSunTIT04}, \cite{MartTrotISIT07}.  The latter channel model may be viewed as being equivalent to an $(a=1,b,\tau)$ DCSW channel model.
We formally define an $(a, b, \tau)$ \emph{streaming code} as a packet-level erasure code that can recover with decoding-delay constraint $\tau$, from all erasure patterns permitted by the $(a,b,\tau)$ DCSW channel model.  It was shown in \cite{BadrPatilKhistiTIT17}, that the rate $R$ achievable by an $(a,b,\tau)$ streaming code is bounded above as per $R \le \frac{\tau+1-a}{\tau+1-a+b} \triangleq R_{opt}$.
This bound is tight as constructions that achieve this bound for all parameter sets $\{a,b,\tau\}$ are presented in \cite{FongKhistiTIT19,NikPVK}. Simulation is typically used to evaluate the performance of the streaming code over the GE channel approximated  by the DCSW channel model, as analytic performance evaluation over the original GE channel is challenging.  We take an important first step here towards analytical performance evaluation.  


\subsubsection*{Notation}
We set $[a,b]=\{a, a+1, \cdots, b\}$, $[b] = [1,b]$, $x^+ = \max\{0, x\}$, $\bar{x} = 1-x$.  For $j \ge i$, $e_i^j = (e_i, e_{i+1},\cdots, e_j)$ and $w(e_i^j) = |\{ i' \in [i,j]\mid e_{i'} = 1 \}|$ and span$(e_i^j) = i_{max}-i_{min}+1$ where $i_{max} = \max\{ i' \in [i,j]\mid e_{i'} \ne 0\}$ and $i_{min} = \min\{ i' \in [i,j]\mid e_{i'} \ne 0\}$. For $j < i$, $e_i^j = \phi$, $w(e_i^j)=0$.  Also, $e_{i_1}^{j_1} \cup e_{i_2}^{j_2} = (e_i \mid i \in [i_1,j_1] \cup [i_2, j_2])$ and $\mathsf{1}^T \triangleq [1 \ 1]$.
\subsubsection*{Outline}
In Section ~\ref{sec:pre} we first provide background on GE channel and streaming codes, then describe our problem setup and contributions.  We present methods to calculate probability of a set of erasure patterns in Section~\ref{sec:comp_prob}.  In Section~\ref{sec:blockbounds} we present the computation of BEP for MDS codes followed by upper bounds on BEP of cyclic codes. In Section~\ref{sec:escbound} we present upper and lower bounds on BEP of ESC codes and conclude in Section~\ref{sec:concl}.
\section{Background} \label{sec:pre}
\subsubsection{GE Channel}
\begin{figure}[ht!]
	\begin{center}
		\includegraphics[width=0.35\textwidth]{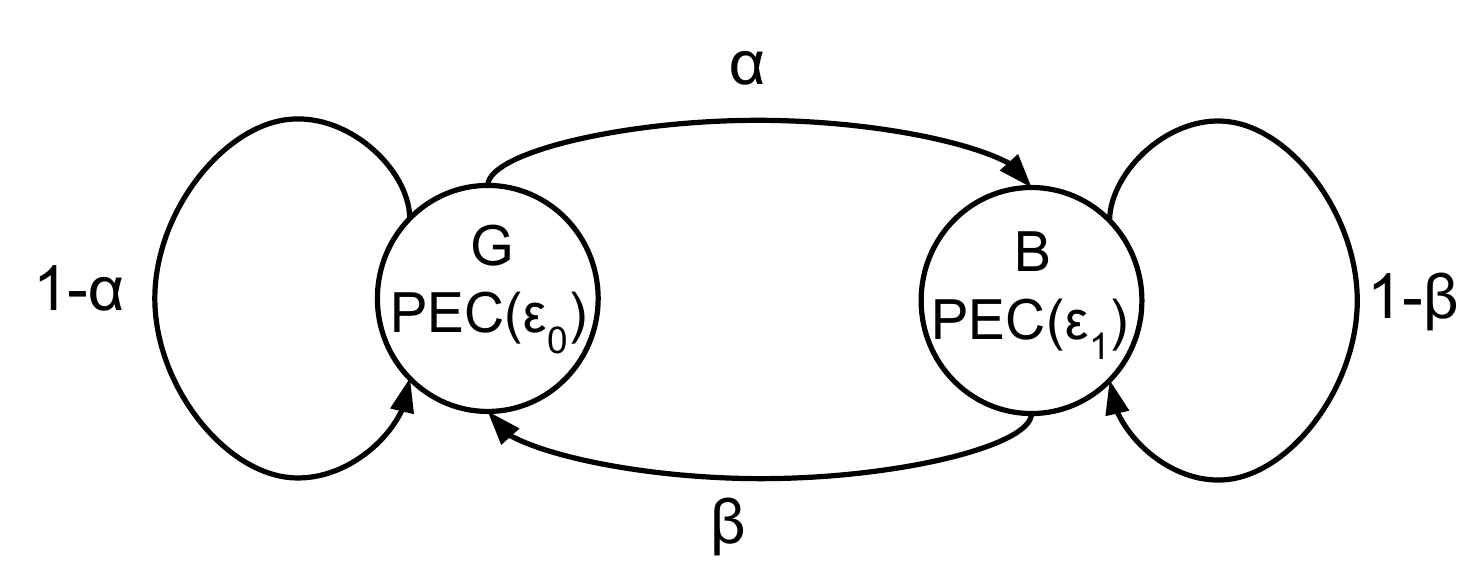}
		\caption{The GE($\alpha, \beta, \epsilon_0, \epsilon_1$) packet-erasure channel model.} \label{fig:GE}
	\end{center}
\vspace{-0.2in}
\end{figure}

The GE channel is a 2-state Markov channel, defined by parameters $(\alpha, \beta, \epsilon_0, \epsilon_1)$. The channel has two states, good and bad, indexed by $0,1$ respectively. Here, in the good state, the channel is PEC($\epsilon_0$) and in bad state it is PEC$(\epsilon_1)$, where PEC$(\epsilon)$ is a packet erasure channel with packet erasure probability  $\epsilon$ 
 and $\alpha, \beta$ are transition probabilities as shown in Fig.~\ref{fig:GE}.
Most of the literature on GE channel focuses on the version of the channel where errors are introduced in place of erasures.  In \cite{YeeWeld}, the authors present an analytical expression for the probability $P(n, k)$ that $k$ out of $n$ symbols transmitted across GE error channel are in error. In \cite{SharHasDho}, the performance of burst-error correcting codes over the GE error channel is studied and recursive expressions are derived for codeword error probability. A generating-series approach for analytically calculating $P(n, k)$ is presented in \cite{CecBla} along with expressions for various burst-error statistics. In \cite{HasHoh_rand_burst}, a performance comparison between random and burst-error correcting codes over the GE error channel is carried out, by coming up with recursive expressions in both cases for the probability of error. Packet loss probability on the GE erasure channel for a class of burst erasure correcting codes called maximally short codes is analyzed in \cite{MartSunTIT04}.

\subsubsection{Streaming Code Construction}
The streaming code constructions in the literature have employed a packet-expansion approach where each coded packet is composed of $n$ symbols of which $k$ are message symbols and the remaining $n-k$ are parity symbols. The $t$-th packet is represented by $$\underline{x}(t)^T = \left[\begin{array}{cc}
\underline{u}(t)^T & \underline{p}(t)^T
\end{array}\right]=(x_1(t), x_2(t), \cdots, x_n(t))$$ where $\underline{u}(t) \in \mathbb{F}_q^k$ represents message and $\underline{p}(t) \in \mathbb{F}_q^{n-k}$ represents parity. Streaming codes are obtained by diagonal embedding (DE) of an $[n,k]$ scalar block code $\calc$, in such a way that the diagonal stream of coded symbols $(x_1(t), x_2(t+1), \cdots, x_n(t+n-1))$ belongs to the underlying $[n,k]$ scalar code $\calc$ (see Fig.~\ref{fig:de}).  DE of $\calc$ will result in an $(a,b,\tau)$ streaming code (see \cite{NikDeepPVK}), if and only if $\calc$ satisfies the following recovery property for every codeword $\underline{c} = (c_1, c_2, \cdots, c_n)$ belonging to \calc:  
\bit 
\item for any $i \in [1,n]$, the code symbol $c_i$ should be recoverable from $\{ c_j \mid j \in [1, i-1] \} \cup \{ c_j \mid j \in [i, \min\{i+\tau,n\}] \setminus E \}$ for any $E \subseteq [i,n]$ such that $i \in E$ and either $|E| \le a$ or $ \max(E) - \min(E) \le b-1$.
\eit 
We declare an erasure pattern $(e_1,\dots,e_n) \in \{0,1\}^n$ as an admissible erasure pattern (AEP) of the $(a,b,\tau)$ DCSW channel if for all $i \in [1,n-\tau]$ either $w(e_i^{i+\tau}) \le a$ or $\text{span}(e_i^{i+\tau}) \le b$.
It can be seen that the above required recovery property of the scalar block code $\calc$ is equivalent to guaranteeing recovery of all code symbols with delay $\tau$ from all AEP of the $(a,b,\tau)$ DCSW channel. We will refer to a scalar block code satisfying this recovery property as an $(a,b,\tau)$ \emph{Embedded Scalar Code} (ESC). Note that any $[n=\tau+1,k=\tau+1-a]$ MDS code satisfies this property for the special case $a=b$ and it's DE results in a rate-optimal $(a,b=a,\tau)$ streaming code. In \cite{FongKhistiTIT19, NikPVK, NikDeepPVK, KhistiExplicitCode,Small}, $[n=\tau+1-a+b,~k=\tau+1-a]$ ESCs are presented for all valid $\{a,b,\tau\}$, giving rise to rate-optimal streaming codes for all parameter sets. While the above description has focused on DE, it is also possible to derive a streaming code through horizontal embedding as discussed below.

\subsection{Problem Setup}

In this paper, we study the erasure-recovery performance of ESCs over the GE channel. 
The performance metric we use here is \emph{block-erasure probability} (BEP). For a block code, we say that a block-erasure has occurred if at least one erased code symbol is not recoverable from non-erased code symbols.  Let $\mathcal{E} \subseteq \{0,1\}^n$ be the set of erasure patterns that can be recovered by an $[n, k]$ block code $\calc$, then the BEP of $\calc$ over GE$(\alpha, \beta, \epsilon_0, \epsilon_1)$ channel is defined as: $$BEP(\calc) =  1- \sum\limits_{e_1^n \in  \mathcal{E}} P(e_1^n),$$
where $P(e_1^n)$ is the probability of seeing erasure pattern $e_1^n$ over  GE$(\alpha, \beta, \epsilon_0, \epsilon_1)$ channel. In the context of ESC, block-erasure means that at least one code symbol is not recovered within the delay constraint. We note that a ESC might correct some erasures outside the admissible patterns of the DCSW channel, but this depends on the exact structure of the scalar code. Since our goal here is to see how well DCSW channel approximates GE channel we assume that an $(a,b,\tau)$ ESC recovers only the admissible erasure patterns of $(a,b,\tau)$ DCSW channel, within delay constraint $\tau$.  The BEP of $(a,b,\tau)$ ESC $\calc_s$ over GE$(\alpha, \beta, \epsilon_0, \epsilon_1)$ channel is thus defined as:
$$BEP(\calc_s) = 1-\sum\limits_{e_1^n \in AEP} P(e_1^n),$$     
where $AEP \subseteq \{0,1\}^n$ is the set of admissible erasure patterns of $(a,b,\tau)$ DCSW channel.
\begin{figure}
	\bc
	\subfigure[DE \label{fig:de}]{\includegraphics[width=0.3\textwidth]{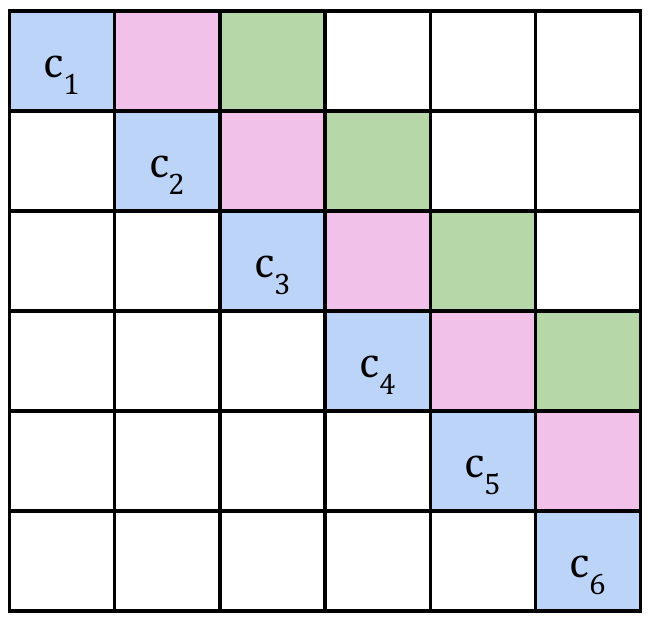}}
	\hspace{0.3in}
	\subfigure[HE of $\ell=6$  codewords\label{fig:he}]{\includegraphics[width=0.3\textwidth]{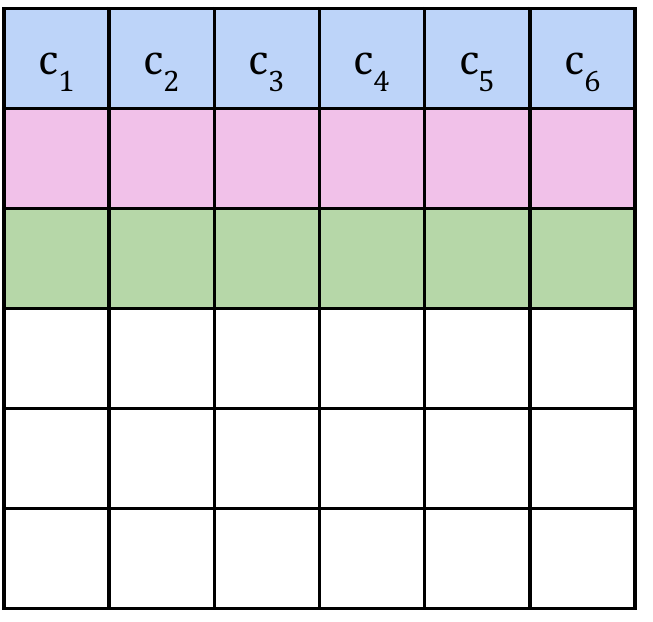}}
	\ec
	\vspace{-0.2in}
	\caption{Embedding of $[6, 4]$ scalar code to get $(a=2, b=3, \tau=4)$ streaming code. Each column is a packet.}
	\vspace{-0.2in}
\end{figure}

\emph{An Application of BEP Analysis}: Streaming codes can also be constructed by introducing separate parity packets. In this method, for every $k$ message packets there will be $(n-k)$ parity packets. Here the idea is to horizontally embed $\ell$ codewords of an $[n,k]$ scalar code to form $n$ packets each containing $\ell$ symbols (See Fig.~\ref{fig:he}). We will repeat this process for every $k$ message packets. 
The packet level code obtained by horizontal embedding (HE) of scalar block code $\mathcal{C}$ is an $(a,b,\tau)$ streaming code if and only if $\mathcal{C}$ is an $(a,b,\tau)$ ESC. Thus, the scalar block codes used in diagonal embedding method to obtain rate-optimal streaming codes can be used here as well to obtain rate-optimal streaming codes. The stream of packets obtained via HE can be divided into blocks each containing $n$ packets, such that that no two blocks share symbols from same codeword.  
Hence the BEP of streaming codes based on HE is equal to BEP of ESC.
\begin{figure}[ht!]
	\bc
	\includegraphics[width=0.5\textwidth]{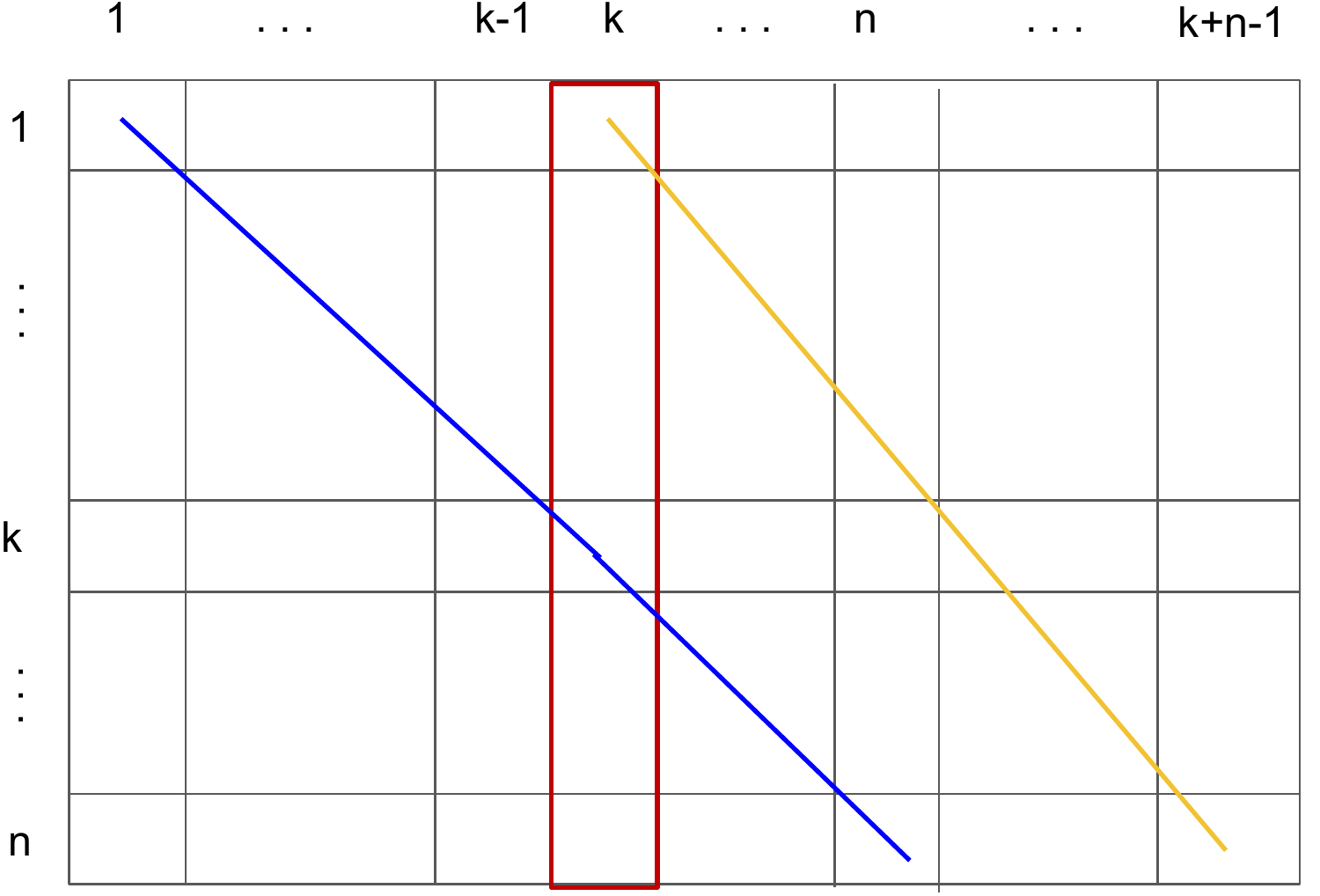}
	\caption{Each column here is a coded packet and row is a symbol within the packet. Say we are interested in recovery of $k$-th packet. The erasures $E_1^n$ impact recovery of message symbol $x_k(k)$ whereas the erasures $E_k^{k+n-1}$ impact recovery of $x_1(k)$. \label{fig:de_pep}}
	\ec
\end{figure}

\emph{Packet Erasure Probability For DE of ESC}: Without loss of generality we assume recovery of $k$-th packet. Packet erasure probability (PEP) for streaming code obtained by DE is given by:
\bean
PEP &=& P\left( (E_k = 1) \cap \left(\cup_{i=1}^k D_i(E_{k-i+1}^{k-i+n}) \right) \right); \\
\eean
where $D_i$ is the event that we fail to recover $x_i(k)$ and this depends on the erasures seen in the interval $[k-i+1:k-i+n]$ given by $E_{k-i+1}^{k-i+n}$ as shown in Fig~\ref{fig:de_pep}.

Note that if $E_{k-i+1}^{k-i+n} \in $ AEP then it implies successful recovery of $x_i(k)$. Therefore the failure event is a subset of  $E_{k-i+1}^{k-i+n} \notin $ AEP. Hence we get:
\bea
\label{eq:aep_bound} PEP &\le& P\left( (E_k = 1) \cap \left(\cup_{i=1}^k (E_{k-i+1}^{k-i+n} \notin AEP) \right) \right)\\
\nonumber &\le& k P(E_1^n \notin AEP) 
\eea 
Note that this is a loose bound and can be improved further by 
\ben
\item characterizing the superset of AEP in equation \eqref{eq:aep_bound} that would still allow for recoverability of $x_i(k)$,
\item by coming up with bounds on $P(E_{i}^{i+n-1} \notin AEP ~|~E_k = 1)$ instead of $P(E_i^{i+n-1} \notin AEP)$ and
\item by tightening the bound for $P\left(\left(\cup_{i=1}^k (E_i^{i+n-1} \notin AEP) \right)|E_k=1 \right)$ instead of using union bound.
\een

\subsubsection*{Our Contributions}
\begin{figure}[ht!]
	\begin{center}
		\includegraphics[width=0.3\textwidth]{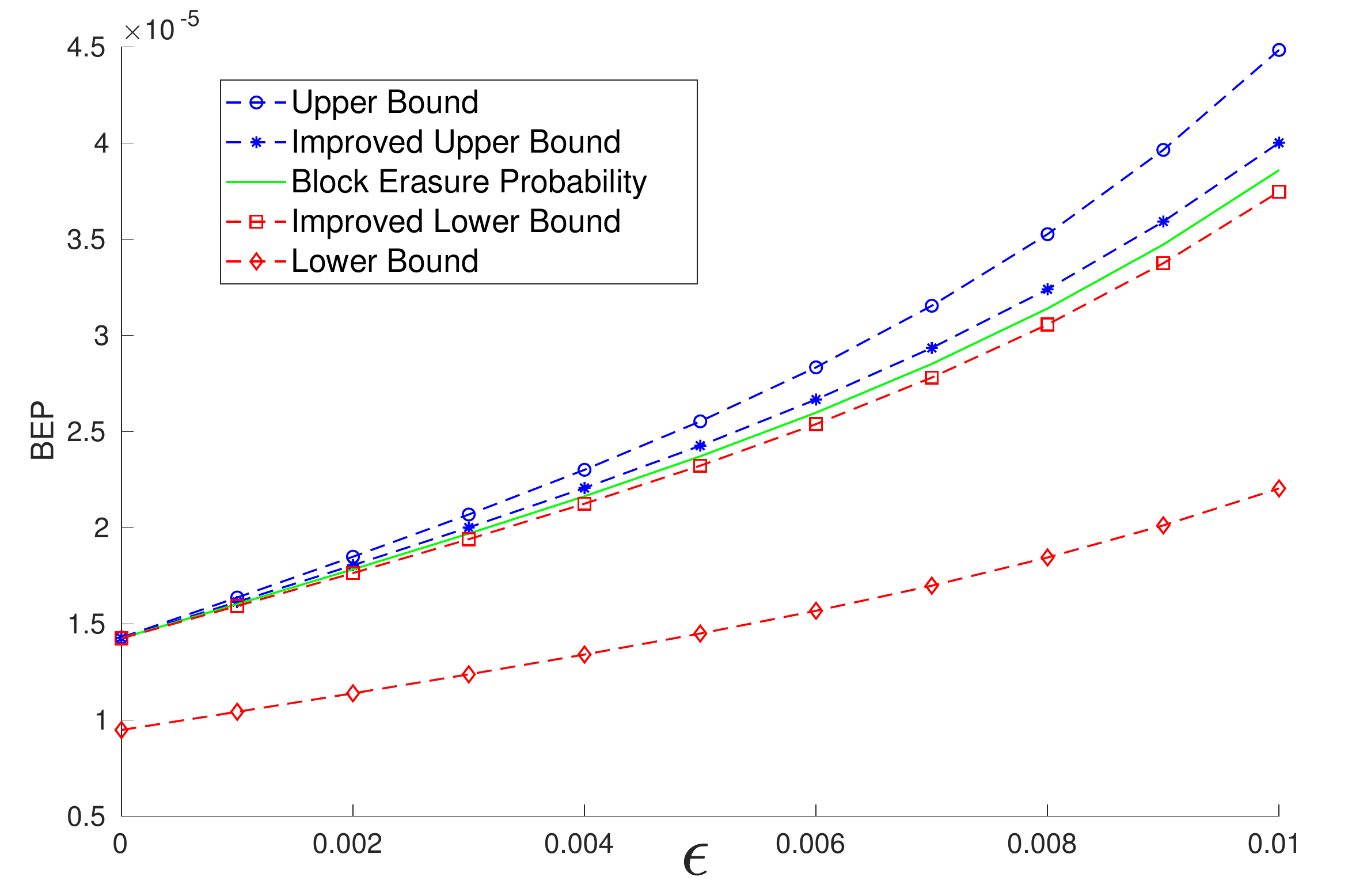}
		\caption{\label{fig:randburstublb} Bounds on BEP for $(a=3, b=6, \tau=10)$ ESC with parameters $(n=14,k=8)$ over GE$(\alpha=10^{-4}, \beta=0.5, \epsilon_0=\epsilon, \epsilon_1=1)$. Exact BEP plot shown above is obtained by enumerating all the AEP and then computing their probabilities.}
	\end{center}
	\vspace{-0.2in}
\end{figure}

 In the present paper, we provide analytical expressions for the BEP over GE channel of block codes that can recover exactly from either at most $a$ random erasures or a burst erasure of size at most $b$. 
 Direct computation of the BEP for ESCs is computationally difficult as it would require enumerating over all admissible erasure patterns. In order to make this tractable, we come up with a superset and a subset of the AEP set whose probability can be computed in polynomial time. This results in upper and lower bounds for BEP of ESCs. We observe that these bounds closely approximate BEP for various $(a,b,\tau)$ values, GE channel parameters (see an example in Fig.\ref{fig:randburstublb}). The upper bounds we prove for BEP allows for picking the best possible $a, b$ for a given decoding delay $\tau$ and BEP requirement $P_{Th}$ (see Fig.~\ref{fig:rateadat}).

\subsubsection*{Outline} In Section~\ref{sec:comp_prob}, we first provide a recap on how to compute probability of seeing $k$ erasures in a window of size $n$ over GE channel. We then follow it up with a window based method to calculate probability of a set of erasure patterns that are characterized by their weights in smaller windows.  In Section~\ref{sec:blockbounds} we present the computation of BEP for MDS codes followed by upper bounds on BEP of cyclic codes. In Section~\ref{sec:escbound} we present upper and lower bounds on BEP of ESC codes and conclude in Section~\ref{sec:concl}.

\section{Computing Probabilities in GE channel\label{sec:comp_prob}}

In this section, we first show computation of probability of seeing $k$ erasures in an $n$ length window, $P(n,k)$.  The closed form expression for this is presented in \cite{CecBla}, we provide a simple proof here for completeness. We will then introduce a window based method to calculate probability of a set of erasure patterns that are characterized by their weights in smaller windows. Let $S_t \in \{0,1\}$ be the random variable denoting the state of GE channel at time $t$ and let $E_t \in \{0,1\}$ be the random variable indicating the presence of erasure at time $t$, where $E_t = 1$ indicates an erasure at time $t$. Thus $P(E_t = 1 | S_t = 0) = \epsilon_0, \  P(E_t = 1 | S_t = 1) = \epsilon_1$. Let $\pi_g = \frac{\beta}{\alpha+\beta}$ denote the steady state probability in good state. For any set $T \subseteq \{0,1\}^n$, any element $e_1^n \in \{0,1\}^n$	we use the notation $P(T)=P(E_1^n \in T)$ and $P(e_1^n) = P(E_1^n = e_1^n)$.
   
    \subsubsection{Probability of seeing $k$ erasures in $n$ instances} 
 
    Let $\pi = P(S_t = 0)$. We first provide a closed form expression for the probability of seeing $k$ erasures in an $n$ length window $[t+1, t+n]$ such that $(t+n)$-th state is $0,1$, given by $g(n,k, \pi)$, $b(n,k, \pi)$ respectively.  More formally, $g(n,k,\pi) = P(w(E_{t+1}^{t+n}) = k, S_{t+n} = 0)$ and 
    $b(n,k,\pi) = P(w(E_{t+1}^{t+n}) = k, S_{t+n} = 1)$. This is clearly independent of $t$ given $\pi$. Probability of seeing $k$ erasures in an $n$-length window $P(n,k)$ can be computed as $P(n,k) = g(n,k,\pi_g)+b(n,k,\pi_g)$.
    
    Given the probability of being in bad state for $r$ out of $n$ instances and ending in good, bad states, given by $g_B(n,r,\pi)=P(w(S_{t+1}^{t+n}) = r, S_{t+n} = 0)$, $b_B(n,r,\pi)=P(w(S_{t+1}^{t+n}) = r, S_{t+n} = 1)$ respectively, the probabilities $g(n,k, \pi)$ and $b(n,k, \pi)$ can be computed as shown below:
%
	\bea
	\label{eq:gbnk}
		\left[\begin{array}{c}
		g(n,k, \pi)\\
		b(n,k, \pi)
	\end{array}\right] = \sum\limits_{r=0}^n \sum\limits_{b=(k+r-n)^+}^{\min\{r, k\}} {r \choose b}{n-r \choose k-b} (\bar{\epsilon}_0)^{n+b-k-r}\epsilon_0^{k-b}(\bar{\epsilon}_1)^{r-b}\epsilon_1^b \left[\begin{array}{c}
		g_B(n,r, \pi)\\
		b_B(n,r, \pi)
	\end{array}\right].
	\eea
    We find the generating functions corresponding to $g_B(n,r, \pi)$ and $b_B(n,r, \pi)$ given by:  $$G_B(L,Z)=\sum\limits_{n=0}^{\infty}\sum\limits_{r=0}^{\infty}g_B(n,r,\pi)L^nZ^r$$ and $$B_B(L,Z)=\sum\limits_{n=0}^{\infty}\sum\limits_{r=0}^{\infty}b_B(n,r,\pi)L^nZ^r$$ respectively.
      \begin{figure}[ht!]
    	\begin{center}
    		\includegraphics[width=0.35\textwidth]{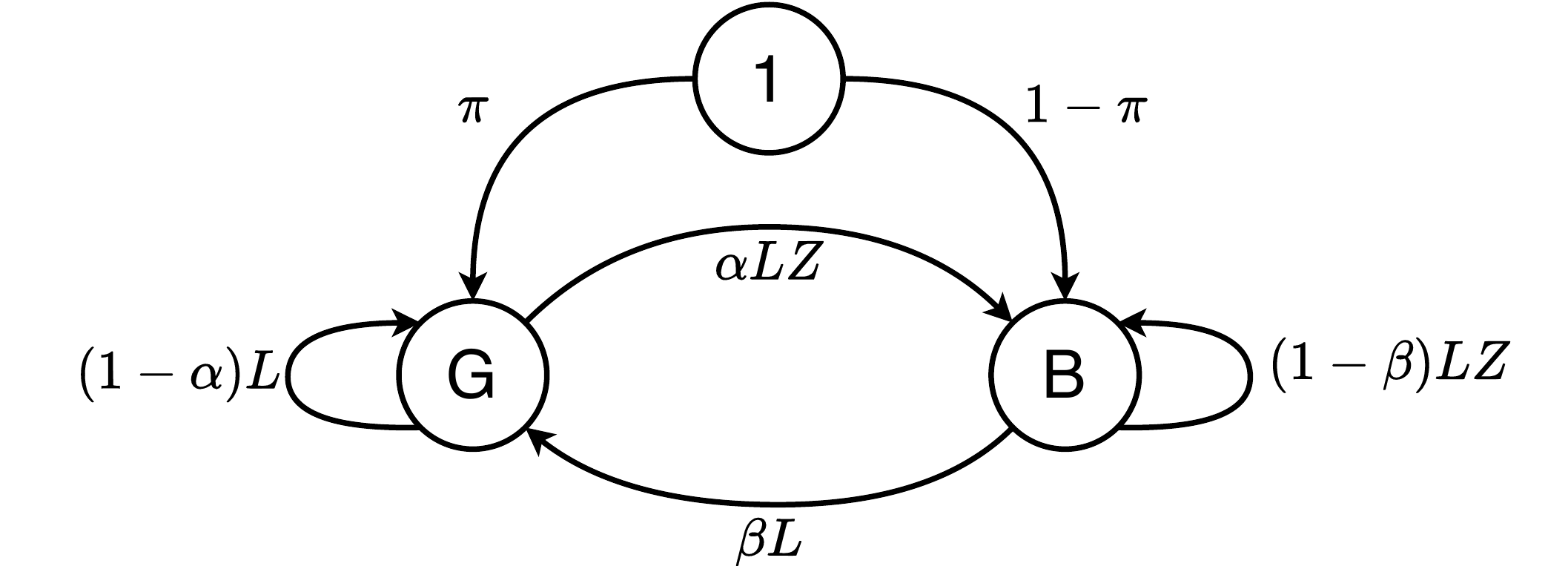}
    		\caption{State diagram illustrating the recursion of generating functions $G_B(L,Z)$ and $B_B(L,Z)$. \label{fig:genfunc} }
    	\end{center}
    	\vspace{-0.17in}
    \end{figure}

      By the state diagram shown in Fig.~\ref{fig:genfunc}, the generating function satisfies following recursion: 
    \bea
  \nonumber  \left[\begin{array}{c}
  	G_B(L,Z)\\
  	B_B(L,Z)
  	\end{array}\right] &=& \left[\begin{array}{cc}
  	\bar{\alpha}L & \beta L\\
  	\alpha LZ & \bar{\beta}LZ
  	\end{array}\right] 	\left[\begin{array}{c}
  	G_B(L,Z)\\
  	B_B(L,Z)
  	\end{array}\right] + \left[\begin{array}{c}\pi \\ \bar{\pi} \end{array}\right]\\
  \label{eq:genfn}&=&  M(L,Z)\left[\begin{array}{cc}
  	1-\bar{\beta}LZ & \beta L\\
  	\alpha LZ & 1-\bar{\alpha}L
  	\end{array}\right] 	 \left[\begin{array}{c}\pi \\ \bar{\pi} \end{array}\right] \\
  \nonumber\text{where } M(L,Z) &=& (1-\bar{\alpha}L-\bar{\beta}LZ+(1-\alpha-\beta)ZL^2)^{-1}\\
  \nonumber &=& \sum\limits_{n=0}^{\infty}\sum\limits_{r=0}^{\infty} m(n,r) L^n Z^r
  \eea

    From \eqref{eq:genfn}, $g_B$ and $b_B$ can be determined as shown below:
	\bea
	\label{eq:gbfromm} \left[\begin{array}{c}
    g_B(n,r, \pi)\\
    b_B(n,r, \pi)
	\end{array}\right] = \left[\begin{array}{ccc}
	\pi & \beta\bar{\pi} & -\bar{\beta} \pi\\
	\bar{\pi} & -\bar{\alpha}\bar{\pi} & \alpha \pi\\
\end{array}\right] \left[\begin{array}{c}
m(n,r)\\
m(n-1,r)\\
m(n-1,r-1)
\end{array}\right]
	\eea
	We will now show a closed form expression for $m(n,r)$. By the power series expansion of $M(L, Z)$ we have:
	\bean
	M(L,Z) &=&  \sum\limits_{a=0}^{\infty} (\bar{\alpha}+\bar{\beta}Z-(1-\alpha-\beta)LZ)^aL^a \\
	&=&  \sum\limits_{a=0}^{\infty} \sum\limits_{z=0}^a \sum\limits_{b=0}^{z} {a \choose z} {z \choose b} L^{a+b} Z^{z} (\alpha-\bar{\beta})^{b} \bar{\beta}^{z-b}  \bar{\alpha}^{a-z} 
	\eean
	In order to obtain $m(n,r)$, we set $z = r$, $b = n-a$ to get:
	\bea
	\label{eq:mnr}  m(n,r) = \sum\limits_{a = \max\{r,n-r\}}^n {a \choose r} {r \choose n-a} (\alpha-\bar{\beta})^{n-a} \bar{\beta}^{a-n+r} \bar{\alpha}^{a-r}.
	\eea
	From \eqref{eq:mnr}, \eqref{eq:gbnk} and \eqref{eq:gbfromm} we have an analytical expression for computation of $g(n,k,\pi), b(n,k,\pi)$. We will now use this to come up with analytical expressions for BEP of random erasure correcting codes, burst erasure correcting codes and either random or burst erasure correcting codes in the next section. 

	\subsubsection{Window based method to calculate probability of erasure patterns}
	In the following Lemma we will show a method to compute the probability of set of erasure patterns defined by their weight in smaller windows.
	\blem\label{lem:window} Let the window $[1,n]$ be partitioned into $\ell$ smaller windows defined by $0 = n_0 < n_1 < n_2 < \cdots < n_{\ell} = n$.  Let $(i_1, i_2, \cdots, i_{\ell})$ be such that $i_j \le n_j - n_{j-1}$ for all $j \in [\ell]$. Then the probability of seeing $i_j$ erasures in window $[n_{j-1}+1, n_j]$ for all $j \in [\ell]$ is given by:
	\bean
	P_e &=& 
1^T Q_{\ell}Q_{\ell-1}\cdots Q_1 \Pi
\eean
where $Q_j = \left[\begin{array}{cc}
	g(n_j-n_{j-1}, i_j, 1) & g(n_j-n_{j-1}, i_j, 0)\\
	b(n_j-n_{j-1}, i_j, 1) & b(n_j-n_{j-1}, i_j, 0)\\
\end{array}\right]$ and $\Pi = \left[\begin{array}{cc}
\pi_g &
\bar{\pi}_g
\end{array}\right]^T$.
	\elem
    \bprf 
    We prove this by induction. For $\ell = 1$ it is clear that: 
    \bean
    \left[\begin{array}{c}
    P(w(E_1^{n_1}) = i_1, S_{n_1} = 0)\\
    P(w(E_1^{n_1}) = i_1, S_{n_1} = 1)
    \end{array}\right] = \left[\begin{array}{cc}
    g(n_1, i_1, 1) & g(n_1, i_1, 0)\\
    b(n_1, i_1, 1) & b(n_1, i_1, 0)
\end{array}\right] \Pi,
    \eean
    by the definition of functions $g, b$.
    Now let us assume that for $\ell = j$:
    \bean
     \left[\begin{array}{c}
    	P(w(E_{n_{j'-1}+1}^{n_{j'}}) = i_{j'}, S_{n_j} = 0, \ \forall j' \in [j])\\
    	P(w(E_{n_{j'-1}+1}^{n_{j'}}) = i_{j'}, S_{n_j} = 1, \ \forall j' \in [j])
    \end{array}\right] = Q_j Q_{j-1} \cdots Q_1 \Pi.
   \eean
   Therefore:
   \bea
   	\nonumber &&
   \left[\begin{array}{c}
   	P(w(E_{n_{j'-1}+1}^{n_{j'}}) = i_{j'}, S_{n_{j+1}} = 0, \ \forall j' \in [j+1])\\
   	P(w(E_{n_{j'-1}+1}^{n_{j'}}) = i_{j'}, S_{n_{j+1}} = 1, \ \forall j' \in [j+1])
   \end{array}\right] \\
	\nonumber &=&
       \sum\limits_{s_{n_j}}\left[\begin{array}{c}
   P(w(E_{n_{j'-1}+1}^{n_{j'}}) = i_{j'}, S_{n_j}=s_{n_j}, S_{n_{j+1}} = 0, \ \forall j' \in [j+1])\\
   P(w(E_{n_{j'-1}+1}^{n_{j'}}) = i_{j'}, S_{n_j}=s_{n_j}, S_{n_{j+1}} = 1, \ \forall j' \in [j+1]) \\
\end{array}\right] \\
\label{eq:winditer}&=&
   Q_{j+1} \left[\begin{array}{c}
   	P(w(E_{n_{j'-1}+1}^{n_{j'}}) = i_{j'}, S_{n_j} = 0, \ \forall j' \in [j])\\
   	P(w(E_{n_{j'-1}+1}^{n_{j'}}) = i_{j'}, S_{n_j} = 1, \ \forall j' \in [j])
   \end{array}\right] \\
\nonumber &=&
    Q_{j+1} Q_j \cdots Q_1 \Pi. 
    \eea
    Equation \eqref{eq:winditer} follows as $Q_{j+1}(s_{n_{j+1}}, s_{n_j}) = P(w(E_{n_j+1}^{n_{j+1}})= i_{j+1}, S_{n_{j+1}} = s_{n_{j+1}} | S_{n_j} = s_{n_j})$. 
    \eprf
    
     The probability of observing an erasure pattern $e_1^n$ can be obtained by setting $\ell = n$, $n_j = j$ for all $j \in [0, n]$ and $(i_1, \cdots, i_n) = (e_1, \cdots, e_n)$ in Lemma~\ref{lem:window}.
    \bean
    P(E_1^n = e_1^n) &=& \mathsf{1}^T \Psi_{e_n} \Psi_{e_{n-1}}\cdots \Psi_{e_1} \Pi, 
    \eean
    where $\Pi =\left[\pi_g, \ \bar{\pi}_g \right]^T$, $\Psi_0 = (I-\Gamma) \Psi$ and $\Psi_1 = \Gamma \Psi$ given
    
    $$\Psi = \left[\begin{array}{cc}
    	1-\alpha & \beta\\
    	\alpha & 1-\beta
    \end{array}\right], \ \ \Gamma = \left[\begin{array}{cc}
    	\epsilon_0 & 0 \\
    0	& \epsilon_1
    \end{array}\right].$$
    

    \section{Block Codes over GE channel\label{sec:blockbounds}}

   The computation of BEP of a block code can be done by characterizing the set $E$ of correctable erasure patterns and then computing the probability of the set $\{0,1\}^n \setminus E$. Characterizing a subset of correctable erasures E results in an upper bound on BEP whereas characterizing a superset of $E$ results in a lower bound.
    
	\subsubsection{Random erasure correcting code}
	The probability of observing erasure patterns with weight larger than $a$ in window of length $n$ over GE$(\alpha, \beta, \epsilon_0, \epsilon_1)$ channel is given by:
	\bea
	\label{eq:prand} P_{rand}(n, a) &=& P( \{e_1^n \mid w(e_1^n) > a\}) \nonumber \\ &=& \sum\limits_{i=a+1}^n g\left(n,i, \pi_g \right)+b\left(n,i, \pi_g \right).
	\eea
	Suppose an $[n, k=n-a]$ MDS code is used over GE channel. Then, the BEP of the MDS code is given by probability of observing erasure patterns of weight larger than $a$. Therefore, BEP of an $[n, k=n-a]$ MDS code is equal to $P_{rand}(n, a)$. 
	\subsubsection{Burst erasure correcting code }
    Here we calculate the probability of erasure patterns whose span is greater than $b$ in a window of length $n$ given by $P_{burst}(n, b)$. It is clear to see that BEP of any $b$ burst erasure correcting code with block length $n$ is upper bounded by $P_{burst}(n ,b)$. Let $B = \{e_1^n \in \{0,1\}^n \mid \text{span}(e_1^n) \le b\}$ be the set of erasures whose span is at most $b$. The set $B$ can be partitioned by the index $i \in [n]$ where the erasure starts, 
  $B = \{\underline{0}\} \cup_{i=1}^n \{ e_1^n \mid e_1^{i-1} = 0, e_i = 1, \text{span}(e_1^n) \le b \}$. 
  The sum of probability of the erasure patterns where the first erasure is at index $i$ is given by:
	\bea
	\nonumber b_i &=& P(\{ e_1^n \mid e_1^{i-1} = 0, e_i = 1, \text{span}(e_1^n) \le b \}) \\  
	\nonumber &=& P(\{ e_1^n \mid e_1^{i-1} = 0, e_i = 1, e_{i+b'}^n = 0 \}), \ b'= \min(b, n-i+1)\\
	 &=& \nonumber  \mathsf{1}^T \sum\limits_{e_{i+b'-1}=0}^{1} \cdots \sum\limits_{e_{i+2}=0}^1 \sum\limits_{e_{i+1}=0}^1 \Psi_0^{n+1-i-b'}\Psi_{e_{i+b'-1}}\cdots \Psi_{e_{i+1}}\Psi_{1}\Psi_{0}^{i-1}\Pi \\
	\label{eq:burstpart0} &=&  \mathsf{1}^T \Psi_0^{n+1-i-b'}\Psi^{b'-1}\Psi_1\Psi_0^{i-1}\Pi\\
	 \label{eq:burstpart} &=& \nonumber \mathsf{1}^T Q_{n+1-i-b'}\Psi^{b'-1}\Psi_{1}Q_{i-1}\Pi~,\\
     \label{eq:Qdefn} &&  \text{ where } Q_j = \left[\begin{array}{cc}
	 	g(j,0,1) & 	g(j,0,0)\\
	 	b(j,0,1) & 	b(j,0,0)
	 \end{array}\right].
	\eea
	Here \eqref{eq:burstpart0} follows as $\Psi_0+\Psi_1=\Psi$. Therefore,
	\bea
	\label{eq:pburst}
  P_{burst}(n,b) = 1- P(B) = 1-g(n,0,\pi_g)-b(n,0,\pi_g)-\sum\limits_{i=1}^{n} b_i.
	\eea

	\begin{note}\normalfont\label{rem:pdontcare}
	In the computation of $P_{burst}(n,b)$, the matrix $\Psi^{i}$ can be computed easily by writing it in the following form:
	\bean
	\Psi^i &=& \left[ \begin{array}{cc}
		1-\mu(1-\rho^i) & (1-\mu)(1-\rho^i)\\
		\mu(1-\rho^i) & 1-(1-\mu)(1-\rho^i)\\
	\end{array} \right],
	\eean	
	where $\rho = 1-\alpha-\beta$ and $\mu = \frac{\alpha}{\alpha+\beta}$. This computation technique    was introduced in \cite{YeeWeld} to analyze the effect of interleaving.
	\end{note}
	\subsubsection{Burst or random erasure correcting code}
	Consider an $[n,k]$ block code $\calc$ that can recover only from erasure patterns given by $E = A \cup B$ where $A$ is the set of erasure patterns that have weight at most $a$ and $B$ is the set of erasure patterns that have span at most $b$. The BEP of this code when used over GE is given by:
	\bea
	\nonumber  P_{rand, burst}(n,a,b)  &=&  1-P(A \cup B) = 1-P(A)-P(B)+P(A \cap B)\\
	\label{eq:randburst} &=&  P_{rand}(n,a)+P_{burst}(n,b)-1+P(A \cap B).
	\eea
	where the last equation follows from equations \eqref{eq:prand} and \eqref{eq:pburst}. The $P(A \cap B)$ can be counted in a way similar to $P_{burst}(n,b)$. The erasure patterns in the set $A \cap B$ can be partitioned into subsets indexed by $i$, where $i \in [n]$ is the index where first erasure is observed. The weight of erasure patterns are limited to be $\le a$ here. For a given $i$, the span of erasure can be atmost $b' = \min(b, n-i+1)$. For a fixed $i$, the probability of these erasure patterns can be determined by dividing the $n$-length window into $\ell = 4$ small windows such that $n_0=0, n_1=i-1, n_2=i, n_3=i+b'-1, n_4=n$ with weights of erasures in each of these windows being $
	(i_1, i_2, i_3, i_4) = (0, 1, <= a-1, 0)$. Then the probability of erasure patterns in the set indexed by $i$ can be determined from Lemma \ref{lem:window} as  $$a_{i} =
	1^T Q_{n-i-b'+1} M \Psi_1 Q_{i-1} \Pi,$$ where $$M =  \sum\limits_{i_3=0}^{\min\{a-1, b'-1\}} \left[\begin{array}{cc}
		g(b'-1, i_3, 1) & g(b'-1, i_3, 0)\\
		b(b'-1, i_3, 1) & b(b'-1, i_3, 0)
     \end{array}\right],$$ and $Q_j$ is defined as shown in \eqref{eq:Qdefn}.
    By also adding the probability of not seeing any erasure in the $n$ length window we get:
	\bea
	\label{eq:paintb}P(A \cap B) &=& g(n,0,\pi_g) + b(n, 0, \pi_g) + \sum\limits_{i = 1}^{n} a_i.
	\eea
	By substituting \eqref{eq:paintb} in \eqref{eq:randburst} we get the analytical expression for BEP of a code that can only correct from either $a$ random erasures or a burst erasure of size atmost $b$. 

    It follows from definition that $$P_{rand,burst}(n, a=1, b) = P_{burst}(n,b)$$ and $$P_{rand,burst}(n, a=b, b) = P_{rand}(n,b).$$ Hence,
    $$P_{rand}(n, b) \le P_{rand, burst}(n, a, b) \le P_{burst}(n, b) \text{ for all } a \in [b].$$ We note that $P_{rand, burst}(n,a,b)$ is an upper bound for the BEP of cyclic codes with parameters $[n,k=n-b, d_{min}=a+1]$. This follows as the cyclic code can correct either $a$ random erasures or a burst erasure of size $b$.	
	
When an $[n,k]$ block code is used, any lost symbol can be recovered, if recoverable, within a decoding delay of $\tau=n-1$. If we compare $[n = \tau+1,k=n-b]$ block codes which can recover only from erasure patterns that have weight atmost $a$ or span atmost $b$, then it can be seen that block code with $a=b$ (MDS code) gives the smallest BEP. However ESCs offer a family of codes that encompass MDS family of codes and can offer better BEPs in comparison to MDS family under same rate and same delay constraint by picking $n > \tau+1$. 


 \section{ESC  over GE channel \label{sec:escbound}}

    Now consider the setting where a ESC that is designed for DCSW channel is used over a GE channel. 
    The BEP can be computed by enumerating all the non admissible erasure patterns and then summing over their probabilities. However this has computational complexity exponential in block length. We therefore propose tractable upper bounds and lower bounds for BEP in this section that can be computed in polynomial time. In the following Lemma we show an upper and lower bound for BEP of $(a,b,\tau)$ ESCs over GE channel using expression for BEP of random or burst erasure correcting block codes. We then improve these bounds in the Lemmas that follow. Let $P_{ESC}(n, a, b, \tau)$ be the BEP of $(a, b, \tau)$ ESC, of block length $n$.
\blem\label{lem:simrandburstbounds}
Let  $n \ge \tau+1$. Then,
$P_{rand,burst}(\tau+1, a, b) \le P_{ESC}(n, a, b, \tau) \le P_{rand,burst}(n, a, b)$.
\elem
\bprf
Let $A_j$ and $B_j$ be defined as
$$A_j = \{e_1^n \mid w(e_j^{\tau+j}) \le a\}, \
B_j = \{e_1^n \mid \text{span}(e_j^{\tau+j}) \le b\}.$$
The erasure patterns that are admissible by $(a,b,\tau)$ DCSW channel are given by $\cap_{j=1}^{n-\tau} (A_j \cup B_j)$. Therefore,
$$P_{ESC}(n, a, b, \tau) = 1-P\left(\cap_{j=1}^{n-\tau} (A_j \cup B_j) \right).$$
By definition $$P_{rand, burst}(n,a,b)=1-P(R_A \cup R_B)$$ where,
$$R_A=\{e_1^n \mid w(e_1^n) \le a \}, \ R_B=\{e_1^n \mid \text{span}(e_1^n) \le b \}.$$
Clearly $R_A\cup R_B$ is a subset of admissible erasure patterns. Therefore the upper bound follows.
By definition in equation \eqref{eq:randburst}, $$P_{rand, burst}(\tau+1, a, b) = 1-P(A_1 \cup B_1).$$ Clearly this is a superset of admissible erasure patterns resulting in the lower bound.
\eprf 

The optimal rate $(a, b, \tau)$ ESCs given in  \cite{NikPVK, FongKhistiTIT19, NikDeepPVK, KhistiExplicitCode} have $n=\tau+1+b-a$. Here, we present improved bounds for $P_{ESC}(n, a, b, \tau)$, for $\tau+1 \le n \le \tau+b$. In order to obtain improved upper bound, we first define  $\hat{U}$ and then show that $\hat{U}$ is a subset of correctable erasure patterns.
\bea
\label{eq:subsetburstrandom} \ \hat{U} &\triangleq& U_A \cup U_B \\    \nonumber U_B&=&\cup_{i=1}^{n-a}\cup_{b'=a+1}^{\min\{n-i+1, b\}} U_{B, i, b'}  \\
\nonumber U_A &=& \{e_1^n \mid w(e_1^n)\le a \} \cup (\cup_{i=1}^{n-\tau-1}\cup_{b'=0}^a U_{A,i,b'}), \\
\nonumber  U_{B,i,b'} &=& \{ e_1^n \mid w(e_1^{i-1})=0, e_i=e_{i+b'-1}=1,  \\ \nonumber  && w(e_{i+b'}^{\min\{i+b'+\tau-a, n\}})=0,  w(e_i^{i+b'-1})>a\}, \\
\nonumber  U_{A, i, b'} &=& 
	\{e_1^n \mid w(e_1^{i-1})=0, w(e_i^{i+b'-1})=b', e_{i+b'}=0, \\ \nonumber &&
 w(e_{i+b'+1}^{\tau+i})=x\le a-b', a-x-b' < w(e_{\tau+i+1}^n) \le b' \}.
\eea
It can be verified that $U_A$ is disjoint from $U_B$, any two sets $U_{B,i_1, b_1}$, $U_{B, i_2, b_2}$ are disjoint given that either $i_1 \ne i_2$ or $b_1 \ne b_2$ and any two sets $U_{A,i_1, b_1}$, $U_{A, i_2, b_2}$ are disjoint given that either $i_1 \ne i_2$ or $b_1 \ne b_2$.

\blem[Improved Upper Bound]\label{lem:randburstimpub} Let  $\tau+1 \le n \le \tau+b$. Then, $$ P_{ESC}(n, a, b, \tau) \le P_{ESC, \hat{U}}(n, a, b, \tau) \le P_{rand, burst}(n, a, b),$$ where $P_{ESC, \hat{U}}(n, a, b, \tau)=1-P(\hat{U})$,  $\hat{U}$ is defined in \eqref{eq:subsetburstrandom}.
\elem
\bprf
Let $e_1^n \in U_A$. If $w(e_1^n)\le a$, then $e_1^n$ is an admissible erasure pattern. Let $e_1^n \in U_{A,i,b'}$ for some $i \le n-\tau$, then $w(e_i^{\tau+i})\le a$ and it is also true that $w(e_1^{\tau+i}) \le a$. This implies $e_1^n \in \cap_{j=1}^i A_j$. Now looking at window $[i+i_0-1, i+i_0-1+\tau]$ for $i_0 \in [b']$:
\bean
w(e_{i+i_0-1}^{i+i_0-1+\tau})&=& w(e_{i+i_0-1}^{i+b'-1})+w(e_{i+b'}^{i+\tau})+w(e_{i+\tau+1}^{i+i_0-1+\tau})\\
&\le& (b'-i_0+1)+x+(i_0-1) \le a
\eean
For window $[j, \tau+j]$ such that $j \ge i+b'$:
\bean
w(e_j^{\tau+j}) &\le& w(e_j^n) = w(e_j^{\tau+i})+w(e_{\tau+1+i}^n)\\
&\le& x+b' \le a
\eean
Therefore it is clear that if $e_1^n \in U_{A,i,b'}$ then $e_1^n \in \cap_{j=1}^{n-\tau} A_j$. Therefore $e_1^n \in \cap_{j=1}^{n-\tau} (A_j \cup B_j)$.
Now let $e_1^n \in U_{B,i,b'}$, then when $i > n-\tau-b'+a$, it implies that span$(e_1^n)=b' \le b$. Therefore it is an acceptable erasure pattern. For the case when $i \le n-\tau-b'+a$ in the window $[j, \tau+j]$ for $j \in [i, i+b'-a]$: span$(e_{j}^{\tau+j}) \le b' \le b$. Therefore $e_1^n \in \cap_{j=1}^{i+b'-a} B_j$. For $j \in [i+b'-a+1, i+b'-1]$:
\bean
w(e_j^{\tau+j}) &=& w(e_j^{i+b'-1})+w(e_{i+b'+\tau-a+1}^{\tau+j})\\
&\le& (i+b'-j)+(j-i-b'+a) = a
\eean 
Therefore $e_1^n \in  \cap_{j=1}^{i+b'-a} B_j \cap \cap_{j=i+b'-a+1}^{i+b'-1} A_j$. For $j \in [i+b', n-\tau]$ we have:
\bean
\text{span}(e_j^{\tau+j}) &\le& \tau+j-(i+b'+\tau-a)\\
&\le& j-i-b'+a\\
&\le& (n-\tau)-b'+a \le b.
\eean
Therefore $e_1^n \in \cap_{j=1}^{n-\tau} (A_j \cup B_j)$. Hence $\hat{U} \subseteq \cap_{j=1}^{n-\tau} (A_j \cup B_j)$ and
\bean
P_{ESC}(n,a,b,\tau) = 1-P(\cap_{j=1}^{n-\tau} (A_j \cup B_j)) \le 1-P(\hat{U}) = P_{ESC, \hat{U}}(n,a, b, \tau).
\eean
 It can be verified that $R_A \subseteq U_A$ and $R_B \setminus R_A \subseteq U_B$ and therefore $R_A \cup R_B \subseteq \hat{U}$. The improvement from the upper bound seen in Lemma~\ref{lem:simrandburstbounds} therefore follows.
\eprf

For obtaining an improved lower bound we come up with a super set of correctable erasure patterns given by,
\bea
\label{eq:supersetburstrandom}\hat{L} &\triangleq& L_0 \cup L_A \cup L_B \\
\nonumber  L_0 &=& \{e_1^n \mid w(e_1^{\tau+1}) = 0\}, \\
\nonumber L_A &=& \cup_{i=1}^{\tau+1} L_{A,i}, \  \ L_B = \cup_{i=1}^{\tau+1-a}L_{B,i},\\ \nonumber
L_{A,i} &=& \{ e_1^n \mid w(e_1^{i-1})=0, e_i=1, \\ \nonumber && w(e_{i+1}^{\tau+1} \cup e_{i+b}^{\min\{i+\tau,n\}}) \le a-1 \},\\ \nonumber
L_{B,i} &=& \{ e_1^n \mid  w(e_{i}^{i+b_m-1}) > a, \\ \nonumber && w(e_1^{i-1})=w(e_{i+b_m}^{\tau+1})=w(e_{i+b}^{\min\{i+\tau,n\}})=0 \}
\eea
where $b_m = \min\{b, \tau+2-i\}$. It can be verified that $L_0$, $L_A$, $L_B$ are mutually disjoint, any two sets $L_{B,i_1}$, $L_{B, i_2}$ are disjoint given that $i_1 \ne i_2$ and any two sets $L_{A,i_1}$, $L_{A, i_2}$ are disjoint given that $i_1 \ne i_2$.

\blem[Improved Lower Bound]\label{lem:randburstimplb} Let $\tau+1 \le n \le \tau+b$. Then,
\bean
 P_{rand, burst}(\tau+1, a, b) \le P_{ESC, \hat{L}}(n,a, b, \tau) \le P_{ESC}(n,a, b, \tau)
\eean
where $P_{ESC, \hat{L}}(n, a,b,\tau)=1-P(\hat{L})$,  $\hat{L}$ is defined in \eqref{eq:supersetburstrandom}.
\elem
\bprf 
We will now show that the set $\hat{L}$ is a super set of the admissible erasure patterns given by the set $\cap_{j=1}^{n-\tau} (A_j \cup B_j)$. Let $e_1^n \in \cap_{j=1}^{n-\tau} (A_j \cup B_j)$, suppose $w(e_1^{\tau+1}) = 0$ then clearly $e_1^n \in L_0$. Otherwise, let $i$ be the first index in $[1, \tau+1]$ where an erasure is seen. Then $w(e_1^{i-1})=0, e_i=1$. But we know that $e_1^n$ is admissible and therefore $e_1^n \in A_1 \cup B_1$. Hence it satisfies one of the following conditions:
\ben
\item $w(e_i^{\tau+1}) \le a$,
\item span$(e_i^{\tau+1}) \le b_m$, w$(e_i^{\tau+1}) > a$ where $b_m = \min\{b, \tau+2-i\}$.
\een
We first look at case 1, where $w(e_i^{\tau+1}) \le a$. For the case when $i > n-b$, $e_1^n$ clearly belongs to the set $L_{A,i}$. For the case when $i \le n-b$, it is necessary that $w(e_i^{\tau+1} \cup e_{i+b}^{\min\{\tau+i, n\}}) \le a$. Suppose $w(e_i^{\tau+1} \cup e_{i+b}^{\min\{\tau+i, n\}}) > a$, then in the window $[j, \tau+j]$ where $j = \min\{i, n-\tau\}$ we can show that:
\bean
w(e_j^{\tau+j}) > a, \text{ span }(e_j^{\tau+j}) > b.
\eean
\bean
w(e_j^{\tau+j})&=& \begin{cases}
	w(e_i^{\tau+i})	 & j = i \text{ i.e., } i \le n-\tau\\
	w(e_{n-\tau}^n) & j = n-\tau \text{ i.e., } i > n-\tau
\end{cases}\\ 
&\ge& \begin{cases}
	w(e_i^{\tau+1} \cup e_{i+b}^{\tau+i}) & j = i \\
	w(e_i^{\tau+1} \cup e_{i+b}^n) & j=n-\tau
\end{cases} = w(e_i^{\tau+1} \cup e_{i+b}^{\min\{\tau+i, n\}}) > a.
\eean
As we know that $w(e_i^{\tau+1}) \le a$ it implies that $w(e_{i+b}^{\min\{\tau+i, n\}}) \ge 1$ implying that there is an erasure at index $\ge i+b$ and at $i$. Therefore, $\text{span} (e_j^{\tau+j}) > b$. This contradicts the condition that $e_1^n \in A_j \cup B_j$. Therefore when $e_1^n$ satisfies case 1, it belongs to $L_{A,i}$.

In case 2, $w(e_i^{\tau+1}) > a$ and span$(e_i^{\tau+1}) \le b_m$. Therefore $w(e_{i+b_m}^{\tau+1})=0$, $w(e_i^{i+b_m-1}) > a$. It is also necessary that $w(e_{i+b}^{\min\{\tau+i,n\}})=0$, 
otherwise the window $[j, j+\tau]$ where $j = \min \{i, n-\tau\}$, will have span$(e_j^{j+\tau}) > b$ and $w(e_j^{j+\tau}) > a$ contradicting that $e_1^n \in A_j \cup B_j$.
Therefore it follows that $\cap_{j=1}^{n-\tau} (A_j \cup B_j) \subseteq \hat{L}$ and: 
\bean
P_{ESC, \hat{L}}(n, a, b, \tau) = 1-P(\hat{L}) \le 1-P(\cap_{j=1}^{n-\tau} (A_j \cup B_j))=P_{ESC}(n,a,b,\tau).
\eean

It is clear to see that $L_0 \cup L_A \subseteq A_1$ and $L_B \subseteq B_1 \setminus A_1$. Therefore $\hat{L} \subseteq A_1 \cup B_1$ and it follows that:
\bean
P_{rand, burst}(\tau+1, a, b) = 1-P(A_1 \cup B_1) \le 1-P(\hat{L}) = P_{ESC, \hat{L}}(n, a, b, \tau).
\eean
\eprf 

We note here that the analytical expression for the improved bounds can be obtained by using the techniques developed in Section \ref{sec:blockbounds} and these expressions are computable in polynomial time.

\subsubsection*{Choosing an $a, b$ for a given delay $\tau$ and BEP Threshold $P_{Th}$} 
\begin{figure}[ht!]
	\begin{center}
		\includegraphics[width=0.45\textwidth]{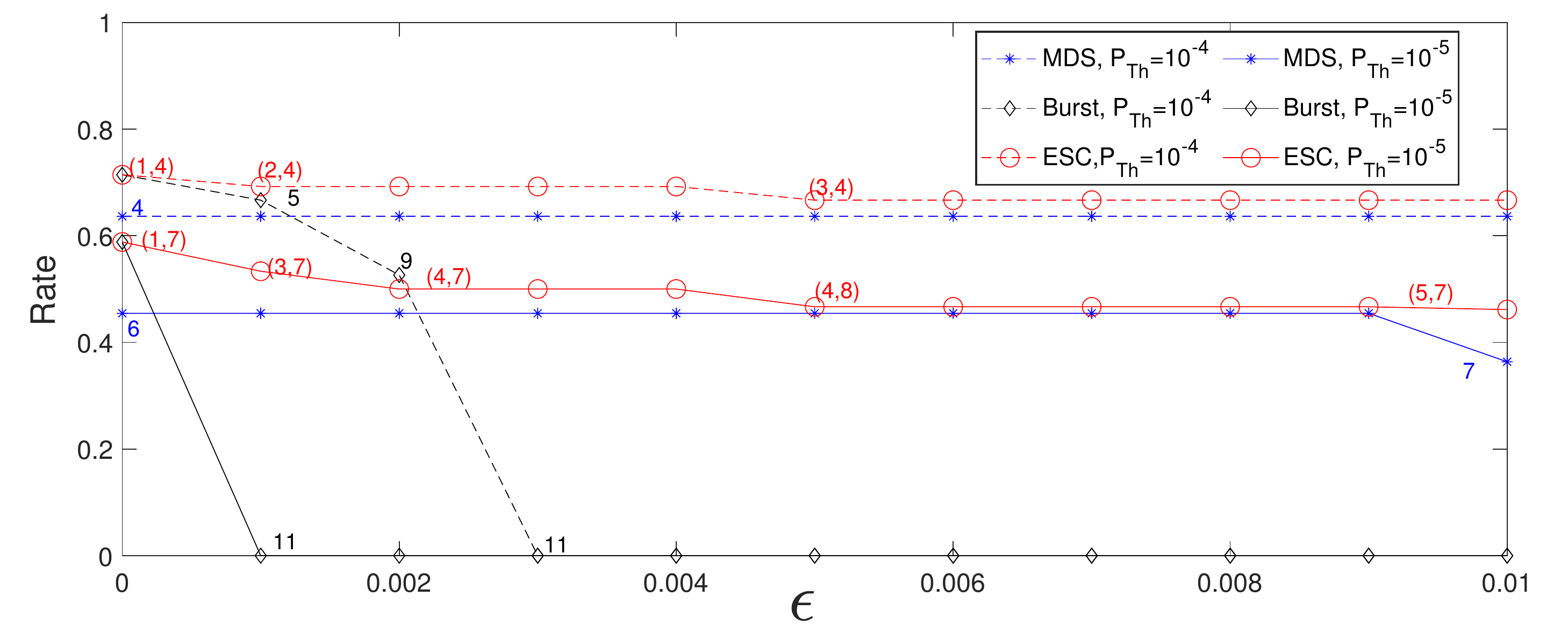}
		\caption{\label{fig:rateadat} The figure depicts rate achievable by MDS code $(b=a)$, burst only ESC $(a=1)$ and ESC when used over  GE$(\alpha=10^{-4}, \beta=0.5, \epsilon_0=\epsilon, \epsilon_1=1)$. Here, decoding delay constraint $\tau=10$, BEP$\le P_{Th}$ and $n=\tau+1+b-a$. The figure also shows code parameters which give the rates shown.}
	\end{center}
	\vspace{-0.2in}
\end{figure}

The parameters $(a, b)$ of $(a, b, \tau)$ ESC can be chosen such that it has highest rate under the constraint $P_{ESC,\hat{U}}(n,a,b,\tau) \le P_{Th}$. This guarantees BEP $\le P_{Th}$ and decoding delay of at most $\tau$.
The figure \ref{fig:rateadat} shows rate gain of ESCs over MDS codes over a GE channel, for the same decoding delay constraint $\tau$ and BEP threshold $P_{Th}$. 
    
%
\section{Conclusion \label{sec:concl}}
In this paper we derived computable upper and lower bounds on BEP of ESCs over GE channel, by characterizing  tractable subset and superset of correctable erasure patterns. This leads to upper and lower bound on BEP of HE based streaming codes. Extending this result to DE based streaming codes by taking into account error propagation remains an open problem.

\newpage 
\bibliographystyle{IEEEtran}
\bibliography{GEAnalysis}		
	
\end{document}